\documentclass[fleqn,usenatbib]{mnras}

\usepackage{newtxtext,newtxmath}

\usepackage[T1]{fontenc}

\DeclareRobustCommand{\VAN}[3]{#2}
\let\VANthebibliography\thebibliography
\def\thebibliography{\DeclareRobustCommand{\VAN}[3]{##3}\VANthebibliography}

\usepackage{graphicx}	
\usepackage{amsmath}	

\title[SN 201015A: observations and analysis]{GRB 201015A: from seconds to months of optical monitoring and supernova discovery}

\author[S. Belkin et al.]{S. Belkin,$^{1,2,3}$\thanks{E-mail: sergey.belkin@monash.edu}
A. S. Pozanenko,$^{1,2,4}$
P. Y. Minaev,$^{2,5}$
N. S. Pankov,$^{1,2}$
A. A. Volnova,$^{2}$
A. Rossi,$^{6}$
G. Stratta, $^{6,7,8}$
\newauthor
S. Benetti,$^{9}$
E. Palazzi,$^{6}$
A. S. Moskvitin,$^{10}$
O. Burhonov,$^{11}$
V. V. Rumyantsev,$^{12}$
E. V. Klunko,$^{13}$
\newauthor
R. Ya. Inasaridze,$^{14}$
I. V. Reva,$^{15}$
V. Kim,$^{15,16}$
M. Jelinek,$^{17}$
D. A. Kann,$^{18}$
A. E. Volvach,$^{12}$
L. N. Volvach,$^{12}$
\newauthor
D. Xu,$^{19}$
Z. Zhu,$^{19}$
S. Fu,$^{19}$
A. A. Mkrtchyan$^{4}$
\\
$^{1}$National Research University “Higher School of Economics”, Myasnitskaya ul. 20, Moscow, 101000 Russia\\
$^{2}$Space Research Institute of the Russian Academy of Sciences, Profsoyuznaya ul. 84/32, Moscow, 117997 Russia\\
$^{3}$School of Physics and Astronomy, Monash University, Clayton, Victoria 3800, Australia\\
$^{4}$Institute of Physics and Technology, Institutskiy Pereulok, 9, Dolgoprudny, 141701, Russia\\
$^{5}$P.N. Lebedev Physical Institute of the Russian Academy of Sciences,  53 Leninsky Avenue, 119991 Moscow, Russia\\
$^{6}$INAF - Osservatorio di Astrofisica e Scienza dello Spazio, via Piero Gobetti 93/3, 40129 Bologna, Italy\\
$^{7}$Institut f\"{u}r Theoretische Physik, Goethe Universit\"{a}t, Max-von-Laue-Str. 1, 60438 Frankfurt am Main, Germany\\
$^{8}$Istituto di Astrofisica e Planetologia Spaziali, via Fosso del Cavaliere 100, I-00133 Roma, Italy\\
$^{9}$INAF – Osservatorio Astronomico di Padova, Vicolo dell’Osservatorio 5, I-35122 Padova, Italy \\
$^{10}$Special Astrophysical Observatory of the Russian Academy of Sciences (SAO RAS), Nizhnij Arkhyz 369167, Russia\\
$^{11}$Ulugh Beg Astronomical Institute (UBAI) of the Uzbek Academy of Sciences, 33 Astronomicheskaya str., Tashkent, 100052, Uzbekistan\\
$^{12}$Crimean Astrophysical Observatory, Russian Academy of Sciences, Nauchnyi, 298409 Russia\\
$^{13}$Institute of Solar-Terrestrial Physics, Russian Academy of Sciences, Siberian Branch, Irkutsk, Russia\\
$^{14}$Evgeni Kharadze Georgian National Astrophysical Observatory, Adigeni, Abastumani, 0301, Georgia\\
$^{15}$Fesenkov Astrophysical Institute, 050020, Observatory street 23, Almaty, Kazakhstan\\
$^{16}$Pulkovo observatory, Russian Academy of Sciences, Saint Petersburg, Russia\\
$^{17}$Instituto de Astrofísica de Andalucíıa (IAA-CSIC), Apartado de Correos, 3.004, E-18.080 Granada, Spain\\
$^{18}$Instituto de Astrofísica de Andalucía (IAA-CSIC), Glorieta de la Astronomía s/n, 18008 Granada, Spain\\
$^{19}$CAS Key Laboratory of Space Astronomy and Technology, National Astronomical Observatories, Chinese Academy of Sciences, Beijing 100101, China\\
}

\date{Accepted XXX. Received YYY; in original form ZZZ}

\pubyear{2015}

\begin{document}
\label{firstpage}
\pagerange{\pageref{firstpage}--\pageref{lastpage}}
\maketitle

\begin{abstract}
We present full photometric coverage and spectroscopic data for soft GRB~201015A with a redshift z = 0.426. Our data spans a time range of 85 days following the detection of GRB. These observations revealed an underlying supernova SN~201015A with a maximum at $8.54\pm$1.48 days (rest frame) and an optical peak absolute magnitude $-19.45_{-0.47}^{+0.85}$ mag. The supernova stands out clearly, since the contribution of the afterglow at this time is not dominant, which made it possible to determine SN’s parameters. A comparison of these parameters reveals that the SN~201015A is the earliest (the minimum $T_{max}$) known supernova associated with gamma-ray bursts. Spectroscopic observations during the supernova decay stage showed broad lines, indicating a large photospheric velocity, and identified this supernova as a type Ic-BL. Thus, the SN~201015A associated with the GRB~201015A becomes the 27th SN/GRB confirmed by both photometric and spectroscopic observations. Using the results of spectral analysis based on the available data of \textit{Fermi}-GBM experiment, the parameters $E_\text{p,i} = 20.0 \pm 8.5$ keV and $E_\text{iso} = (1.1 \pm 0.2) \times 10^{50}$ erg were obtained. According to the position of the burst on the $E_\text{p,i}$--$E_\text{iso}$ correlation, GRB 201015A was classified as a Type II (long) gamma-ray burst, which was also confirmed by the $T_\text{90,i}$--$EH$ diagram.
\end{abstract}




\section{Introduction}

In recent years, extensive optical observations of gamma-ray bursts (GRBs) have revealed a physical relationship between long-duration (T$_{90}\gtrsim$2 s, see~\citet{Kouveliotou93,Koshut96}) GRBs and core-collapse supernovae (SNe)~\citep[e.g.][]{Cano2017Guide}. The first reliable evidence of such an association was found between GRB 980425 and SN 1998bw. Type Ic-BL SN 1998bw was determined to be temporally and spatially consistent with GRB~980425~\citep[][]{Galama98,Iwamoto98,Kulkarni98}. Then, the very bright event GRB~030329 was confirmed to be associated with SN~2003dh~\citep[][]{Hjorth03,Stanek03,Matheson03} which was also classified as type Ic-BL. As of today, there is no unequivocal distinction between SN Ic and SN Ic-BL. It is considered that broad-lined supernovae (SN Ic-BL) are SNe Ic with a higher photospheric velocity approaching 15000 -- 30000 km s$^{-1}$ and absence of the H and He in the spectra~\citep[e.g.][]{Modjaz2016}. In comparison with ordinary SNe, the kinetic energy of these two mentioned events exceeded 10$^{52}$ erg. Such highly energetic SNe were named hypernovae~\citep[e.g.][]{Paczynski98}.


As a result of numerous observations, it is possible to indicatively divide the optical light curve of long GRB into four successively alternating phases~\citep[e.g.][]{Pozanenko2021}. The prompt emission phase can barely be observed due to the relatively slow response of the ground-based observatories to the GRB detection alerts.

The prompt emission phase is followed by the afterglow stage. It is usually the longest one and can be described by a power law or a power law with a jet-break (see, e.g. \citealt{Sari1999}). At this stage, inhomogeneities, characterized by a statistically significant deviation of data points from the power-law decay of the afterglow, can occur~\citep[e.g.][]{Mazaeva2018}. The light curve of the early GRB~201015A afterglow indicates the presence of a bump with a peak at $\sim$6.6 minutes post trigger that deviates substantially from the power-law slope of the afterglow. There are a number of models to describe this behaviour. See~\citet{Ror23} for details.

Further, the bumps identified in the light curves of GRB about 7 -- 30 days after the burst were interpreted as a SN explosion with an expansion of its photosphere, followed by an increase in luminosity~\citep[e.g.][]{bloom1999,garnavich2003}. The radiation of a SN manifests itself when the central engine's activity has long ceased, and there is nowhere to take an additional powerful source of energy. Spectroscopic observations showed that spectra measured during the SN phase are usually characterized by broad lines and the absence of hydrogen and helium lines, which is one of the hallmarks of Type Ic-BL supernovae. See Section~\ref{supernova} for details related to the supernova associated with the GRB~201015A.

Finally, the source fades, causing the afterglow flux of the GRB, and the SN radiation to drop below the luminosity level of the host galaxy. This is the final phase of the optical light curve. The study of this stage makes it possible to estimate both the integrated extinction within the host galaxy and its physical parameters.

This work is dedicated to a comprehensive study and description of the GRB~201015A and associated supernova discovered both in photometric and spectroscopic data. Section~\ref{observations} contains the description of the observations that led to the reported data. Information about the detection of prompt gamma-ray emission and its classification is presented in Section~\ref{prompt_phase}. Section~\ref{afterglow} is dedicated to the the both early and late-time afterglow and its analysis. A discussion of the results of afterglow observations in the X-ray range is also given. The description and analysis of photometric and spectroscopic observations of a SN associated with a GRB~201015A are given in Section~\ref{supernova}. Information about the host galaxy of the GRB~201015A is presented in Section~\ref{host_galaxy}. All results of observations, processing, and interpretation are discussed in Section~\ref{DataAnalysisAndresults} and summarized in Section~\ref{conclusion}.


\section{Observations}
\label{observations}
At 22:50:13 UT (=T$_0$ hereafter) on 2020, October 15 \citep{gcn28632}, GRB 201015A was initially detected by the Burst Alert Telescope (BAT) \citep{Barthelmy2005} aboard the \textit{Neil Gehrels Swift Observatory} (henceforth \textit{Swift}) with duration of $T_\text{90} $ $\approx$ 10 s and relatively soft energy spectrum \citep{gcn28658}. \textit{Swift} was not able to slew immediately to the burst due to an observing constraint. Later, the burst was also found in data of \textit{Fermi}-GBM in a ground-based search of sub-threshold events \citep{gcn28663}.


\subsection{X-ray Imaging}
\label{X-ray Imaging}
The X-ray afterglow was detected and observed by the \textit{Swift}'s X-ray telescope (XRT, \citet{Burrows2005}). XRT began to observe the field of GRB 201015A only 3214.1 seconds after the BAT trigger \citep{gcn28635}. Using XRT data and 4 Ultraviolet/Optical Telescope (UVOT, \citet{Roming2005}) images, an astrometrically corrected X-ray position was determined as following RA(J2000)=23:37:16.96, Dec(J2000)=+53:24:52.6 with an uncertainty of 3.8 arcsec (90\% confidence radius) \citep{gcn28635}.

Consequently, 7200 seconds of XRT data for GRB~201015A from 3200 to 57300 seconds after the BAT trigger were obtained. The light curve in this time interval is preliminarily described with a single power law with a decay index $\alpha = 1.49_{-0.21}^{+0.24}$ \citep{gcn28660}.

\begin{table*}
	\centering
	\caption{The main parameters of GRB 201015A.}
	\label{GRB_parameters}
	\begin{tabular}{lcc}
		\hline
	    Parameter & Value & Reference \\
		\hline
		RA(J2000) & 23h 37m 16.41s & \citet{gcn28637} \\
		Dec(J2000) & +53d 24$^\prime$ 56.5$^{\prime\prime}$ & \citet{gcn28637} \\
        z & 0.426 & \citet{gcn28649} \\
        D$_L^1$ & 2363.6 Mpc & \citet{Wright2006} \\
        A$_R$ & 0.375 & \citet{Schlafly2011} \\
        T$_{90,i}$(Fermi-GBM) & 11.5$\pm$2.5 s & On-line Fermi catalogue\\
        E$_{iso,}$ & (1.1$\pm$0.2)$\times$10$^{50}$ erg & On-line Fermi catalogue; this work\\
        E$_{p,i}$ & 20.0$\pm$8.5 keV & On-line Fermi catalogue; this work\\
        Offset & 1.5$^{\prime\prime}$(8.453 kpc) & this work\\
        \hline
		\multicolumn{3}{l}{$^{1}$ -- Calculated using $H_0 = 69.6$ km s$^{-1}$ Mpc$^{-1}$, $\Omega_M = 0.286$, $\Omega_{\Lambda} = 0.714$ \citep[][]{Bennett2014}.}
	\end{tabular}
\end{table*}

\subsection{Optical Observations}
\label{optical_observations}
Optical observations of the field of GRB~201015A were carried out with the ZTSh telescope~\citep{Rumyantsev19} of Crimean Astrophysical Observatory~\citep[CrAO,][]{Severny1955}, the AZT-33IK telescope of the Sayan observatory~\citep[Mondy,][]{Chuprakov18}, the AS-32 telescope of Abastumani Astrophysical Observatory~\citep[AbAO,][]{Khetsuriani67}, the Zeiss-1000 telescope of Special Astrophysical Observatory of the Russian Academy of Sciences~\citep[SAO RAS,][]{Shvedova95}, the Zeiss-1000 telescope of Tien Shan Astronomical Observatory~\citep[TSHAO,][]{Elenin2015}, the AZT-22 telescope of Maidanak Observatory~\citep[MAO,][]{Ehgamberdiev00}, the AZT-20 telescope of Assy-Turgen Observatory~\citep[ATO,][]{Serebryanskiy18}, the LBT telescope~\citep{Hill94} of Mount Graham International Observatory~\citep[MGIO,][]{Sage03}, the FRAM~\citep{Prouza10} and GTC~\citep{RodEsp97} telescopes of Roque de los Muchachos Observatory~\citep[ORM,][]{Ardeberg84}, the NEXT telescope of Xinjiang Astronomy Observatory (XAO) and the 2.16m telescope~\citep{Zhou16} of the Xinglong Observatory (XO) in a sufficient number of imaging epochs. The characteristics of the telescopes used are presented in Table~\ref{telescopes}.

The earliest observations were made using the 25cm FRAM-ORM robotic telescope at La Palma (Spain) which automatically responded to the alert regarding GRB~201015A \citep{gcn28632} and obtained a series of unfiltered images with an exposure of 20 s, starting at 22:50:50.8 UT, i. e. 37.6 s. since trigger (see~\citet{gcn28664,Ror23}).

Observations with the 2.6m ZTSh telescope of CrAO observatory between Oct. 16 (UT) 01:02:14 and 02:52:23 ($\sim$2.3 to $\sim$4 hours post burst trigger) were carried out in $White$ and $R$ filters. The optical counterpart was clearly detected in each frame of 120 s exposure in the $R$ filter \citep{gcn28656}. The magnitude at this epoch was $R \sim 20.7$ mag and showed a power-law decay.

Subsequent observations of the optical afterglow were conducted using the 1.6m AZT-33IK telescope of Mondy observatory, the 0.7m AS-32 telescope of Abastumani observatory, the 1.5m AZT-20 telescope of Assy-Turgen observatory, and the 1m Zeiss-1000 telescope of SAO RAS of the GRB-IKI follow-up network (FuN). Altogether these observations spanned the time range from 2.3 hours to 85 days after the GRB~201015A detection.

Spectroscopic observations of the GRB 201015A were carried out with the Multi-Object Double Spectrographs MODS-1 and MODS-2 instruments \citep{Pogge2010a} mounted on the 2$\times$8.4-m LBT telescope (Mt Graham, AZ, USA) at the mid-time of 04:00 UT on 2020-11-13, $\sim$28.8 days after the burst trigger~\citep{gcn29306}.

Observations were also conducted on the facilities of other groups \citep[][]{gcn28633,gcn28637,gcn28639,gcn28645,gcn28653,gcn28664,gcn28674,gcn28676,gcn28677,gcn28680,gcn28681}.

\begin{figure}
	\includegraphics[width=\columnwidth]{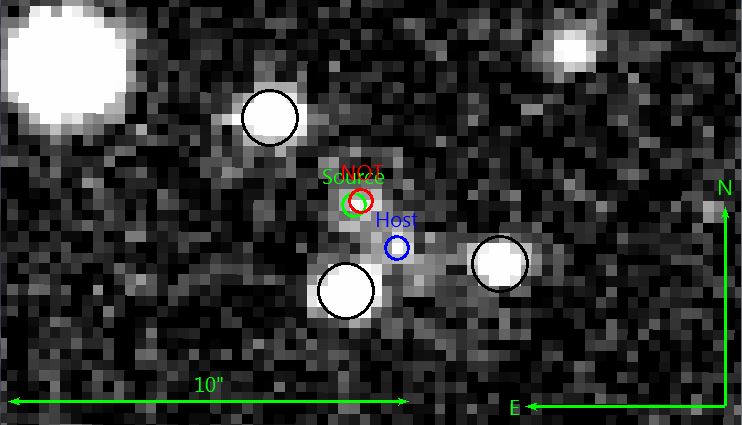}
    \caption{The finding chart image of GRB~201015A. The source is denoted by a green circle. The localisation of the source determined from the results of observations with the NOT \citep{gcn28637}, is represented by a red circle. A blue circle marks the position of the proposed host galaxy. The black circles indicate nearby sources.}
    \label{nearobjects}
\end{figure}

Analysis of the acquired images reveals the presence of several objects in the immediate vicinity of the source (Fig.~\ref{nearobjects}). As a result, the most significant contribution to the resulting light curve was made by observations with telescopes featuring the smallest pixel scales (e.g. AZT-22, GTC, LBT). Nonetheless, this does not diminish the contribution of observations at other observatories.

\begin{table*}
	\centering
	\caption{The telescopes used to observe the burst in the optical band.}
	\label{telescopes}
	\begin{tabular}{lcccc}
		\hline
		Observatory & Location & Telescope & Diameter$^{1}$ & Field of view\\
		\hline
		CrAO & Nauchny/Crimea,   & ZTSh & 2.6 & 14$_\cdot^{\prime}$4$\times$14$_\cdot^{\prime}$4\\
		Mondy & Sayan,   & AZT-33IK & 1.6 & 6$_\cdot^{\prime}$7$\times$7$_\cdot^{\prime}$9\\
		AbAO & Abastumani, Georgia & AS-32 & 0.7 & 44$_\cdot^{\prime}$4$\times$44$_\cdot^{\prime}$4\\
		SAO RAS & Arhyz,   & Zeiss-1000 (S) & 1.0 & 7$_\cdot^{\prime}$05$\times$7$_\cdot^{\prime}$36\\
		TSHAO & Tien-Shan,   & Zeiss-1000 (T) & 1.0 & 19$_\cdot^{\prime}$2$\times$19$_\cdot^{\prime}$2\\
		MAO & Maidanak,   & AZT-22 & 1.5 & 18$_\cdot^{\prime}$3$\times$18$_\cdot^{\prime}$3\\
		ATO & Assy-Turgen,   & AZT-20 & 1.5 & 22$_\cdot^{\prime}$3$\times$22$_\cdot^{\prime}$3\\
		MGIO & Arizona,   & LBT & 8.6 & 6$_\cdot^{\prime}$0$\times$6$_\cdot^{\prime}$0\\
		ORM & The Canary Islands,   & GTC & 10.4 & 8$_\cdot^{\prime}$0$\times$9$_\cdot^{\prime}$7\\
		XAO & Xinjiang,   & NEXT & 0.6 & 21$_\cdot^{\prime}$6$\times$21$_\cdot^{\prime}$8\\
		XO & Xinglong,   & Xinglong-2.16m & 2.16 & 9$_\cdot^{\prime}$4$\times$9$_\cdot^{\prime}$4\\
		\hline
		\multicolumn{5}{l}{$^{1}$ -- The diameter (size) of the primary mirror, m.}
	\end{tabular}
\end{table*}

\subsection{Radio observations}
The observations at 36.8 GHz were carried out with the 22-m RT-22 radio telescope at the foot of Mount Koshka (Simeiz, Crimea). The antenna beam width at half maximum is 100$^{\prime\prime}$. The radio telescope was pointed toward the source alternately by one and the other beam lobes forming under diagram modulation and having mutually orthogonal polarizations.

The antenna temperature from the source was determined as the difference between the radiometer responses averaged over 30 s at two different antenna positions. A series of 200–250 measurements were carried out, whereupon the average signal was calculated and its root-mean-square error was estimated. The measured antenna temperatures corrected for the absorption of emission in the Earth’s atmosphere were converted to flux densities by comparing them with the results of observations of calibration sources \citep{Volvach2023}. The results obtained from the observational data are presented in Table~\ref{radio_obs_log} and shown in Fig.~\ref{FullLC}.

\subsection{Redshift determination}
Spectroscopic observations of the afterglow of GRB 201015A were carried out with the 10.4m GTC Telescope, at Roque de los Muchachos Observatory (La Palma, Spain) equipped with OSIRIS. It was conducted on 2020 Oct 16.172 UT ($\sim$5.28 hours after the GRB detection) and consisted of 3$\times$900 seconds exposures with the R1000B grism, covering the wavelength range from 370 to 780 nm. The spectrum showed a featureless continuum throughout the full spectral range. However, it was possible to identify emission features superposed on the trace, which were identified as [OIII], [OII], and H-beta at a common redshift of 0.426~\citep{gcn28649}.

The obtained redshift value was confirmed as a result of spectroscopic observations with the Nordic Optical Telescope, equipped with the ALFOSC spectrograph. The observations started on 2020 Oct 15.978 UT (38 min after the GRB detection) and consisted of 4$\times$900 seconds exposures with grism, covering the wavelength range from 380 to 940 nm. 

Despite the trace visible across the full spectral range, weak absorption features were detected in a blue part of it and can be interpreted as Mg II, Mg I, and Ca II at the redshift of 0.423~\citep{gcn28661}. Considering faint emission lines such as [O II] and Halpha also allows to confirm determined redshift. It sets a firm lower limit to the GRB's redshift, excluding a chance of superposition with a background galaxy.

\section{Data analysis and Results}
\label{DataAnalysisAndresults}
\subsection{Prompt Emission (gamma-rays)}
\label{prompt_phase}

\begin{figure}
	\includegraphics[width=\columnwidth]{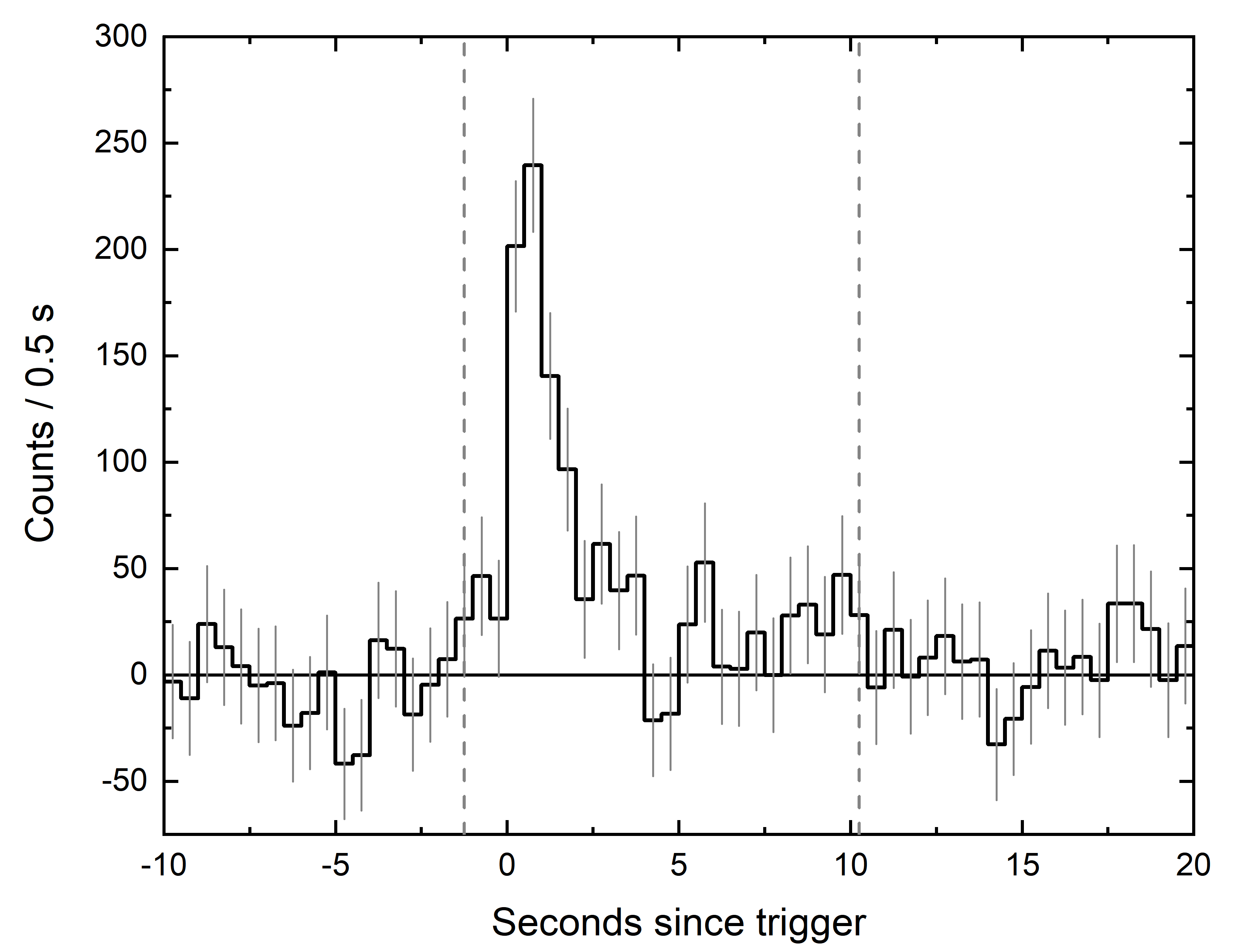}
    \caption{Background subtracted light curve of GRB 201015A in the energy range of (6, 50) keV with a time resolution of 0.5 s, based on \textit{Fermi}-GBM data. Vertical dashed lines indicate $ T_\text{90} $ time interval ($ T_\text{90} $ = 11.5 $\pm$ 2.5 s).}
    \label{fig:gbmlc}
\end{figure}

We use publicly available photon-by-photon (TTE) data of \textit{Fermi}-GBM experiment\footnote{\url{ftp://legacy.gsfc.nasa.gov/fermi/data/}} to analyze temporal properties of GRB 201015A. Background subtracted light curve of GRB 201015A, constructed using data of most illuminated NaI\_10 and NaI\_11 detectors in the energy range of (6, 50) keV, is presented in Fig.~\ref{fig:gbmlc}. The burst is not visible in the energy range above 50 keV due to its soft spectrum. The light curve consists of the main emission episode with a duration of about 2 s, followed by a weak tail with a total duration of $T_\text{90} $ = 11.5 $\pm$ 2.5 s (Fig.~\ref{fig:gbmlc}). The behavior of the light curve is typical for type II (long) bursts, but the classification as type I (short) burst with extended emission is also possible  \citep[e.g.][]{Norris06,Minaev2010,minaev10b,mozg21}.

\begin{figure}
	\includegraphics[width=\columnwidth]{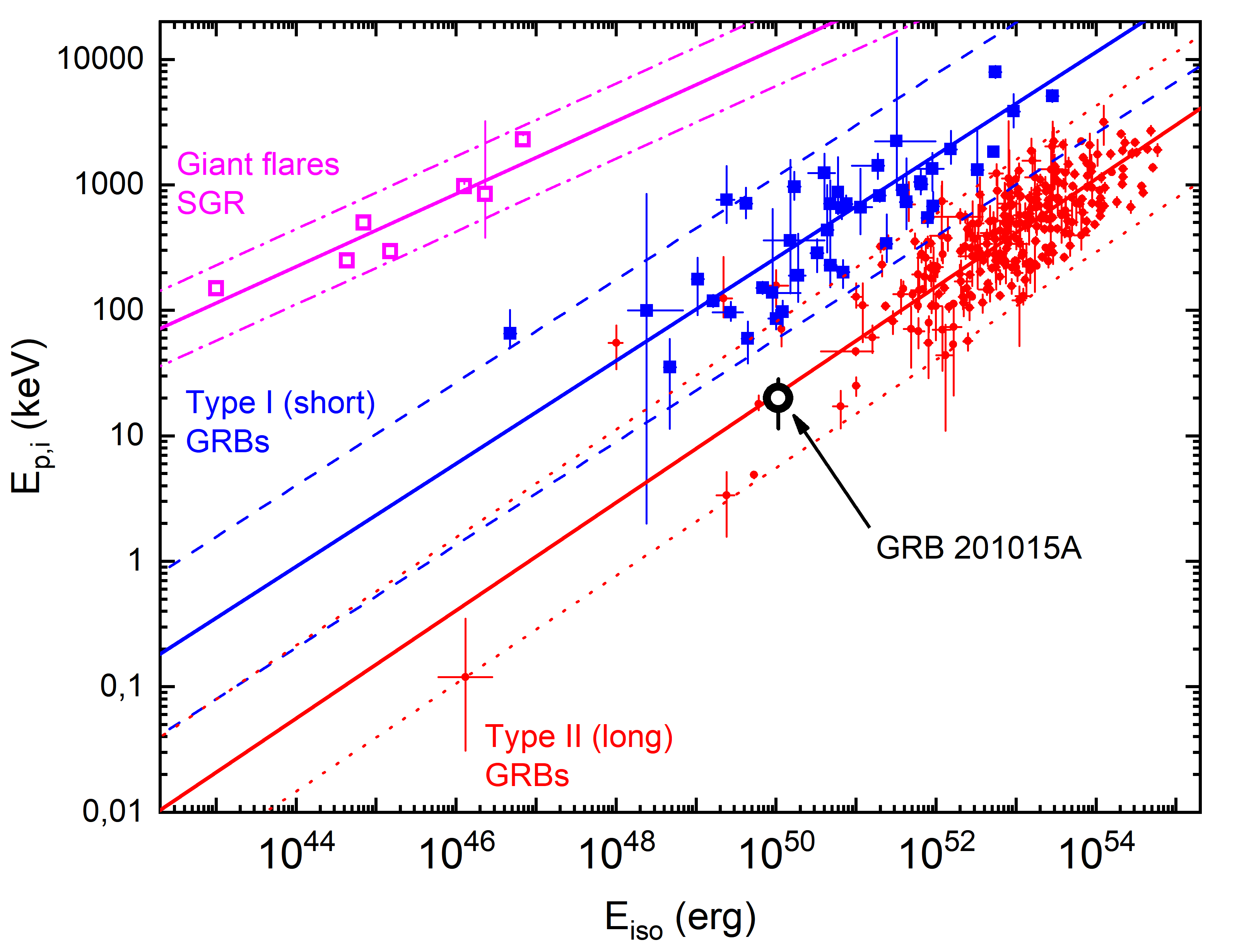}
    \caption{The $ E_\text{p,i} $ -- $ E_\text{iso} $ correlation of for type I (blue squares), type II (red circles) GRBs and SGR giant flares (magenta unfilled squares) with the approximation results, including 2$\sigma_\text{cor}$ correlation regions. The position of GRB 201015A (unfilled black circle) is typical for type II GRBs.}
    \label{fig:amati}
\end{figure}

To classify GRB 201015A, we can use the method proposed in \citet{min20a}, based on $ E_\text{p,i} $ -- $ E_\text{iso} $  correlation \citep[e.g.][]{ama02}, as we know redshift of the source: $z$ = 0.426 \citep{gcn28649}. Using results of GRB 201015A spectral analysis, published in \citet{gcn28663}, we obtain $ E_\text{p,i} $ = 20.0 $\pm$ 8.5 keV and $ E_\text{iso} $ = (1.1 $\pm$ 0.2)$\times$10$^{50}$ erg. The position of GRB 201015A on $ E_\text{p,i} $ -- $ E_\text{iso} $ diagram is shown at Fig.~\ref{fig:amati}, along with samples and correlation analysis results from \citet{min20a,min20b,min21}. The burst is placed close to the $ E_\text{p,i} $ -- $ E_\text{iso} $  correlation fit for type II (long) bursts, characterizing GRB 201015A as type II GRB \citep{gcn28668}. 

The position of the burst at $ E_\text{p,i} $ -- $ E_\text{iso} $ correlation could be characterized by parameter $EH =  \frac{(E_\text{p,i} / 100~\text{keV})}{ (E_\text{iso} / 10^{51}~\text{erg})^{~0.4}}$, firstly introduced in \citet{min20a}. Type I bursts are harder (higher value of $ E_\text{p,i} $) and fainter (lower value of $ E_\text{iso} $) than type II ones in general, therefore they are characterized by a higher value of $EH$ parameter. For GRB 201015A we obtain $EH$ = 0.49, which is close to the most probable value of the type II bursts sample, $EH$ = 0.69 (type I GRBs have $EH$ > 3.3, in general). 

\begin{figure}
	\includegraphics[width=\columnwidth]{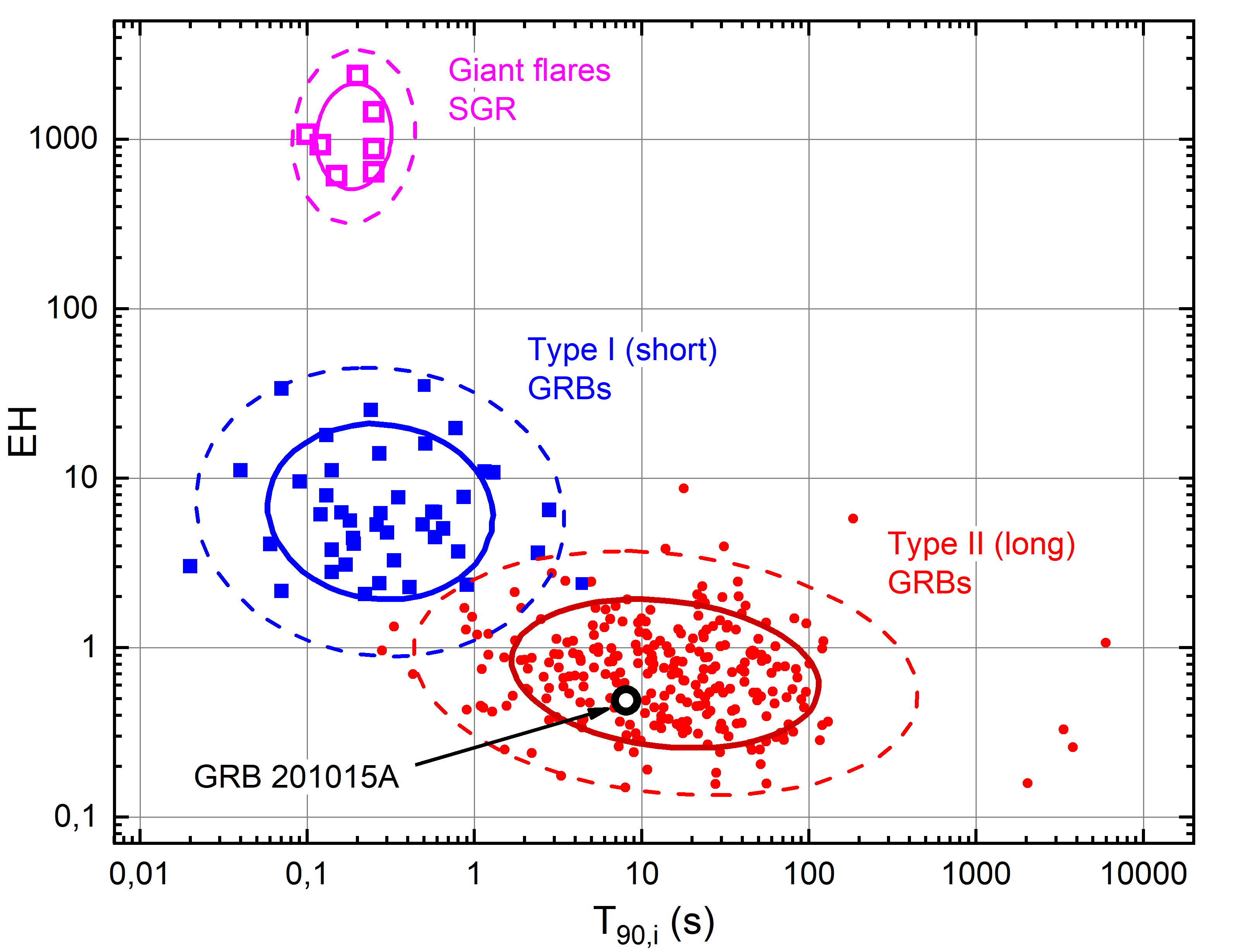}
    \caption{The $ T_\text{90,i} $ -- $ EH $ diagram for type I (blue squares), type II (red circles) GRBs and SGR giant flares (magenta unfilled squares) with corresponding cluster analysis results (68\% and 95\% confidence regions are shown by bold solid and thin dashed curves). GRB 201015A (black unfilled circle) is within the 68\% confidence region of type II gamma-ray bursts.}
    \label{fig:ehd}
\end{figure}

The best possible separation of type I GRBs from type II ones is achieved in the coordinates of $EH$ and $ T_\text{90,i} $ (duration in rest frame), which is shown in Fig.~\ref{fig:ehd}. The $ T_\text{90,i} $ -- $ EH $ diagram clearly classifies GRB 201015A as type II GRB, placed inside 68\% cluster region of type II bursts.

\subsection{Afterglow}
\label{afterglow}

\subsubsection{X-ray afterglow}
The afterglow in the X-ray range was plotted on the basis of data that are publicly available on the \textit{Swift} Burst Analyzer website\footnote{\url{https://www.swift.ac.uk/burst_analyser/}}. It consists of the data within $\sim$0.03 to $\sim$0.3 days time range and also includes a dot $\sim$20 days after the GRB~201015A trigger.

The available data does not allow to confirm the presence of any inhomogeneities since there are no X-ray observations of the light curve between the first and twentieth days. A joint power-law fit of all points gives a slope of 0.76$\pm$0.03.

Late-time observations were carried out by Chandra X-ray Observatory 8.4 and 13.6 days after the trigger \citep{gcn28822}. The power-law decay between these two epochs was found to be $\sim$ -0.8. This value is consistent with the \textit{Swift}'s data approximation, which may also indicate the absence of any inhomogeneities in the X-ray light curve of the afterglow of the GRB~201015A.

Chandra's observations were conducted in the 0.5-7 keV range. The unabsorbed flux values, recalculated for the energy range of 0.3-10 keV, which corresponds to \textit{Swift}-XRT, are given by \citet{Giarratana2022}. It can be seen that these points align well with the extrapolation of the \textit{Swift}-XRT data spanning a time interval from 0.03 to 0.3 days. It confirms the absence of significant inhomogeneities in the X-ray afterglow. Therefore, this slope can be used to describe the optical afterglow.

Altogether X-ray data are shown in Fig.~\ref{FullLC}.

\subsubsection{Optical afterglow}
\label{optical_afterglow}
The data acquired during the optical observations were subjected to a standard preliminary reduction, including the rejection of defective frames, the reduction of frames using a bias matrix, a dark frame, a flat field, and fringe removal, if necessary. the data reduction was performed with the IRAF\footnote{IRAF (Image Reduction and Analysis Facility), an environment for image reduction and analysis, was developed and maintained by the National Optical Astronomy Observatory (NOAO, Tucson, USA) operated by the Association of Universities for Research in Astronomy (AURA) under cooperative agreement with the National Science Foundation of the USA, see \url{iraf.noao.edu}.} software package. The preliminary data reduction was carried out by the \textit{ccdproc} code of the package \textit{ccdred}; the individual images in the corresponding filters were then stacked by the \textit{imcombine} code to provide the best S/N. Aperture photometry was performed with the \textit{apphot} code of the \textit{daophot} package.

\begin{figure*}
	\includegraphics[width=0.9\textwidth]{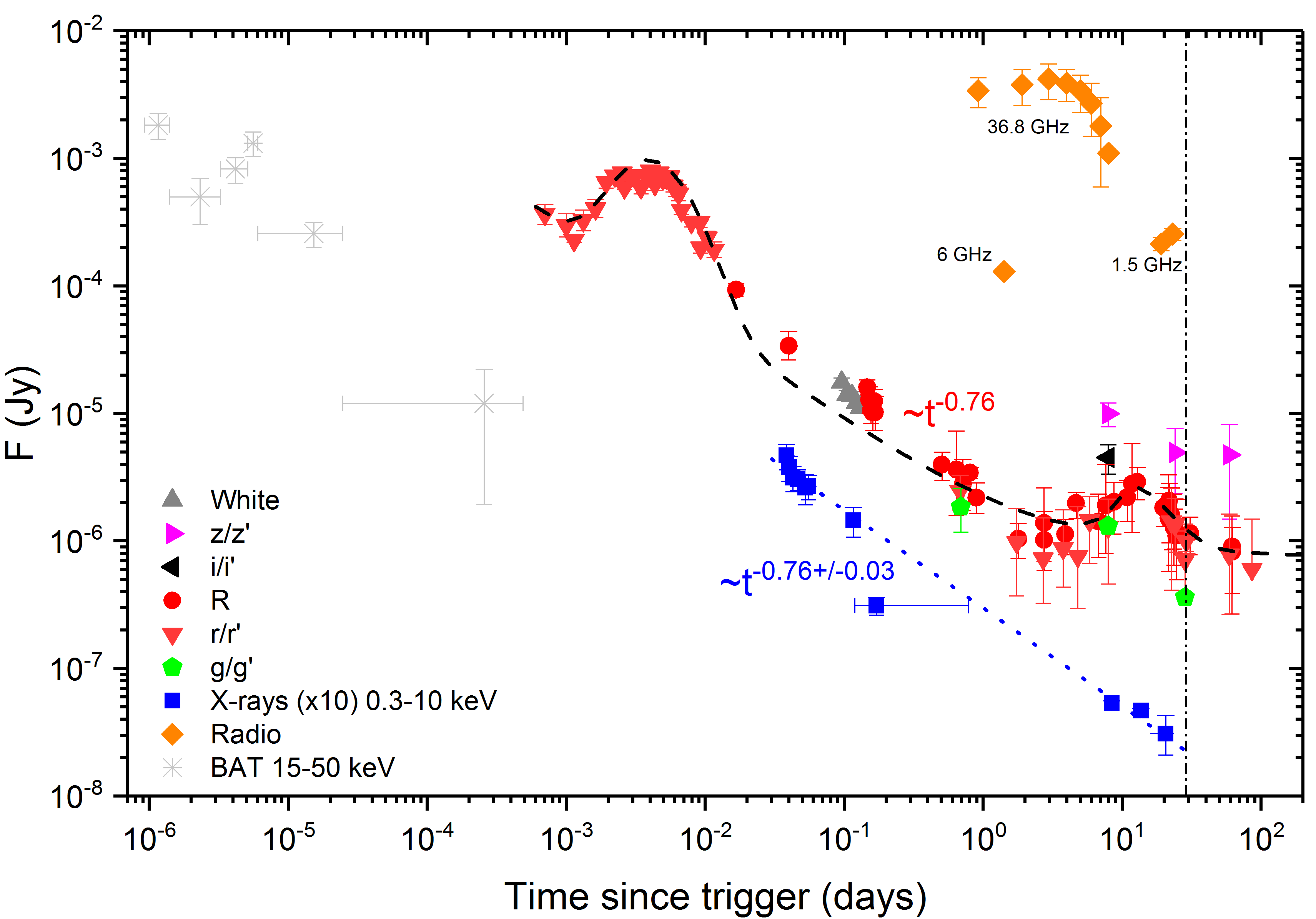}
    \caption{The complete multiwavelength light curve of GRB 201015A. All optical data are shown in Table~\ref{obs_log}. Radio observations at 36.8 GHz are shown in Table~\ref{radio_obs_log}, while those at 1.5 GHz and 6 GHz were taken from \citet{gcn28945} and \citet{gcn28688} respectively. The blue dotted line denotes an approximation of the \textit{Swift}-XRT data. The black dashed line shows the joint approximation of the inhomogeneity at the stage of early afterglow, late afterglow, supernova, and host galaxy (see Section~\ref{SNLC}). The data is not corrected for Galaxy extinction. The vertical dash-dotted line indicates the epoch when the spectrum was taken (see Section~\ref{SpectrumSection}).}
    \label{FullLC}
\end{figure*}

Obtaining a photometric solution for all epochs, from all instruments, and in all filters was achieved through mathematical flux subtraction. Given the close proximity of the source to several other objects simultaneously, direct isolation was barely possible. In this context, the method of template subtraction from scientific images did not allow for an accurate measurement of the source's magnitude due to observations being conducted with a wide array of different instruments. This variability arose from differences in pixel scales, which in most cases exceeded 0.7$^{\prime\prime}$/pixel, as well as variations in the limiting magnitudes depending on the instrument used. Consequently, image subtraction exhibited non-uniformity across all instruments. To mitigate this photometry inhomogeneity, the method of mathematical flux subtraction was employed. The following steps were covered. Aperture photometry was performed, with the radius of the aperture selected to encompass both the source, its host galaxy, and neighboring blending objects, thereby capturing the total flux from all these sources. Throughout the observation period, only the flux of the GRB~201015A and SN~201015A associated with it displayed variations, while the luminosity of the host galaxy remained assumed to be constant. Measurements of magnitudes from nearby sources in GTC and Maidanak images showed no variability. Therefore, during the processing of each observation epoch, the contributions of neighboring objects were subtracted, accounting for the corresponding error transfer, and subsequently, the flux from the GRB~201015A and its host galaxy was determined. The results of this approach are depicted in the light curve presented in Fig.~\ref{FullLC}. The same approach was applied to the host galaxy's magnitude determination.

LBT/MODS imaging observations were taken in the Sloan filters $g^\prime$ and $r^\prime$ under good seeing conditions ($0\farcs6$) and were reduced using the data reduction pipeline developed at INAF - Osservatorio Astronomico di Roma \citep{Fontana2014a} which includes bias subtraction and flat-fielding, bad pixel and cosmic ray masking, astrometric calibration, and coaddition.

The measured instrumental magnitudes were calibrated against the SDSS–DR12 and USNO-B1.0 photometric catalogues. The reference stars were chosen according to the algorithm described in~\citet{Skvortsov2016}.

All optical observations are shown in Table~\ref{obs_log} and encompass a time range of 60 seconds to 85 days after the GRB trigger. It allows to construct the light curve shown in Fig.~\ref{FullLC}. It shows a significant inhomogeneity at the early afterglow stage in the form of a bump. Object brightness peaks about 5 minutes after the GRB trigger with R$\sim$16.5 mag and then fades until the end of the current data set $\sim$1 hour after the trigger. Based on these observations, it was suggested that the inhomogeneity might be related more to the onset of the afterglow than to the reverse shock(see~\citet{Ror23}).

After flare activity, the afterglow follows a single power law decay. The slope was adopted from the X-ray range since no inhomogeneities are observed in it over a given time period. It was confirmed with several different instruments mentioned in Section~\ref{optical_observations}. It is equal to -0.76 and this slope will be used during the joint approximation of early afterglow bump, afterglow, and supernova components of the GRB 201015A's light curve.

Based on this approach, a second inhomogeneity becomes evident, occurring from 30 minutes to 7.2 hours. It manifests itself as a positive deviation of the observation points from the afterglow approximation. However, due to data limitations, this cannot be either confirmed or refuted.

The values of the afterglow approximation parameters can be found in Table~\ref{optic_lc_fit_parameters}.

\subsubsection{Afterglow numerical modelling}
\label{afterglowpy}
We attempt to model the multi-wavelength afterglow of GRB~201015A using a Bayesian inference approach through a MCMC code that explores the parameter space of synthetic afterglow realisations obtained with the publicly available Python package \texttt{afterglowpy}~\citep{Ryan2020}. \texttt{Afterglowpy} is a Python package for modelling GRB afterglows which computes synchrotron radiation from an external shock, allowing to take into account  possible effects due to a jet structure and an off-axis observer position.

We model the full radio to X-ray dataset of GRB~201015A by assuming a Gaussian model of the jet structure and fixing some of the parameters: an electron energy distribution power law index of p=2.2, the fraction of energy that goes into electrons and magnetic field of $\epsilon_e=0.1$ and $\epsilon_B=0.001$, respectively. We find best model of other parameters, with a 50\% percentile (plus 84\% and minus 16\%) kinetic energy of $E_k=2.0^{+3.8}_{-1.35}\times10^{53}$ erg, a density of the surrounding ISM of $n=2\times10^{-3}$ cm$^{-3}$, a jet half opening angle of the inner core of $\theta_j =0.30^{+0.08}_{-0.12}$ rad, and a viewing angle of the observer with respect to the jet axis of $\theta_v=0.31^{+0.30}_{-0.21}$ rad. The optical and X-ray data are well fitted by synchrotron radiation with a cooling frequency well above the X-ray range at the time the supernova component is rising above the afterglow. This scenario is in line with the assumption of modeling the optical flux decay with the same rate measured in X-rays.

We note that this solution does not fit the light curve at 36.8 GHz which is an order of magnitude brighter than the
modeled peak value. This  might indicate the presence of an additional  emission component (e.g. a contribution from reverse shock), that we could not model in  the framework  of \texttt{afterglowpy}.


\begin{figure}
	\includegraphics[width=\columnwidth]{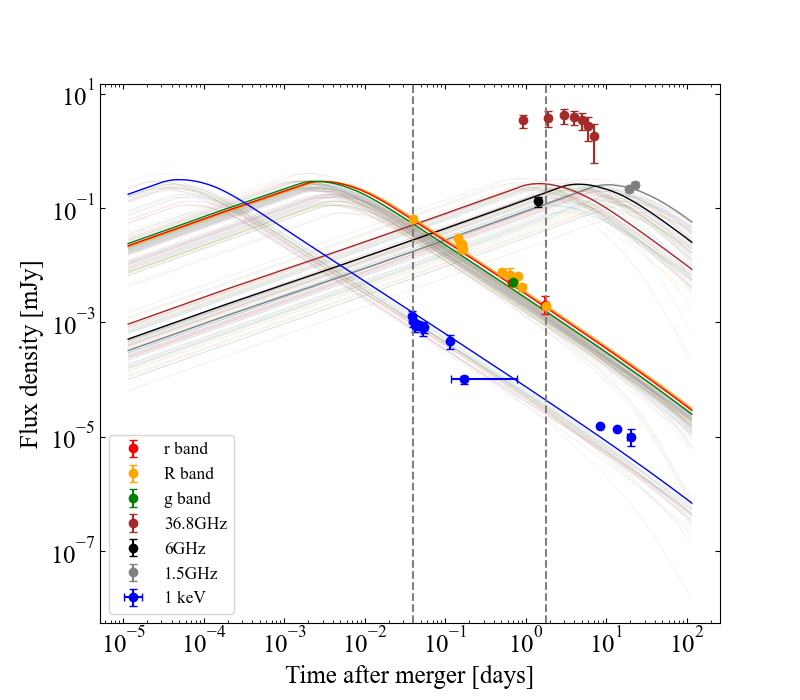}
    \caption{Multi-wavelength afterglow light curves (see Section~\ref{afterglowpy}). For each band, the light curves predicted by the standard model are shown: r, R, and g-bands (red, yellow, and green respectively), 36.8 GHz (brown), 6 GHz (black), 1.5 GHz (grey), X-ray light curve (blue colour). The vertical lines pinpoint the epochs of the SEDs, illustrated in the Figure~\ref{afterglowpy_spec}.}
    \label{afterglowpy_lc}
\end{figure}

\begin{figure}
	\includegraphics[width=\columnwidth]{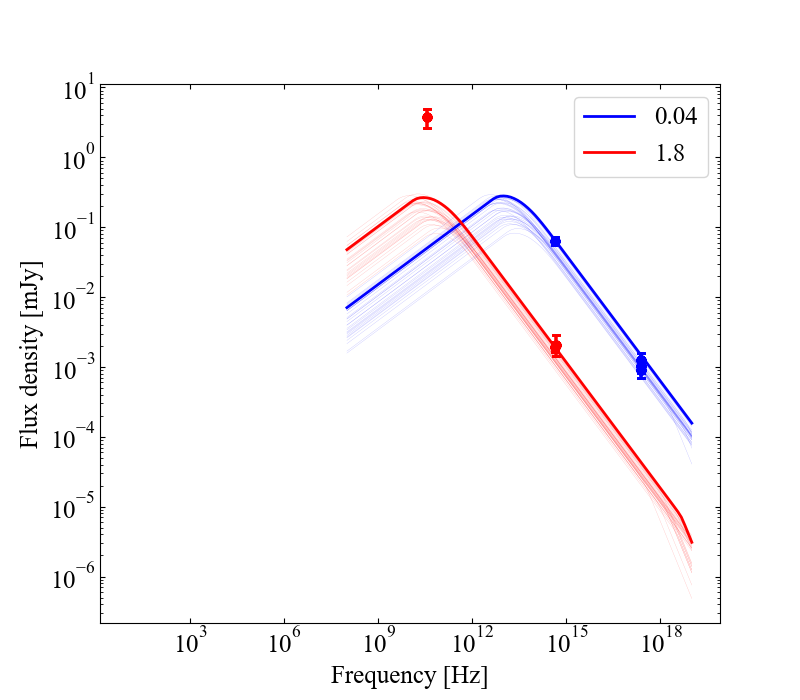}
    \caption{Spectral energy distribution of the GRB~201015A at epochs 0.04 and 1.8 days after detection. The results of the modeling are indicated by lines (see Section~\ref{afterglowpy}).}
    \label{afterglowpy_spec}
\end{figure}

\subsection{Supernova}
\label{supernova}
The supernova associated with GRB~201015A was detected in late photometric observations as an evident bump above the power law afterglow decay~\citep{gcn29033} and confirmed by spectroscopic observation~\citep{gcn29306}.


\subsubsection{Light curve}
\label{SNLC}
Following the standard procedure, the detection of the supernova can be accomplished by subtracting the contribution of the host galaxy and the afterglow from the observed optical light. 

Since most of the observations were made in $r$ and $R$ filters, it was decided to reduce them to a single light curve. Using quasi-synchronous observations in the R filter at the Mondy, CrAO, and AbAO observatories and in the r filter at the Assy-Turgen observatory within the time span of $\sim$0.6 -- 0.8 days after the GRB detection, the shift between these filters was determined. The points in the R filter were approximated, and the value was calculated at an epoch of 0.68081 days when we had observations in the r filter. By assuming that at this stage the source's luminosity decline is achromatic, we subtract the magnitude in the R filter from the magnitude in the r filter. From this procedure, the shift was determined. The photometric values taken with the $r$ filter need to be brightened by 0.514 mag to match the $R$ filter data.

The host galaxy's level was obtained based on late-time observations resulting in a determination of 24.5 mag in the r$^{\prime}$ filter. This contribution has been mathematically subtracted from the optical light curve.

The extinction in the Galaxy in the direction of the source is known ($A_R = 0.735$; based on~\citet{Schlafly2011}) and has also been taken into account.

In order to distinguish the SN component the estimation of the afterglow contribution should be done assuming that it is produced by synchrotron radiation in a relativistic shock interacting with a homogeneous external medium, as predicted by the classical fireball model~\citep[e.g.][]{Sari1998,Zhang2004}.

Since no sufficient inhomogeneities are present in XRT afterglow during the first $\sim$8 days, its slope was utilised to approximate the late-time optical afterglow ($\sim$t$^{-0.76}$) whilst the early afterglow cannot be characterised by a single power law due to the presence of sufficient inhomogeneity (see~\citet{Ror23}). Consequently, an approximation function was formulated to describe several parts of the optical light curve after host galaxy subtraction and Galaxy extinction correction simultaneously (Eq.\ref{eq:OT}):

\begin{equation}
    F_\nu^{OT}(t)=F_\nu^{eAG}(t)+F_\nu^{AG}(t)+F_\nu^{SN}(t)
    \label{eq:OT}
\end{equation}

The individual components, namely the early afterglow bump, the afterglow itself, and the supernova were approximated by the Eq.\ref{eq:eAG}, Eq.\ref{eq:AG}, and Eq.\ref{eq:SN}, respectively:

\begin{equation}
    F_\nu^{eAG}(t)=\frac{A}{\omega_1 x \sqrt{2 \pi}}\exp{-\frac{(\ln{\frac{x}{x_C}})^2}{2 \omega_1^2}}
    \label{eq:eAG}
\end{equation}

\begin{equation}
    F_\nu^{AG}(t)=B x^{-\alpha}
    \label{eq:AG}
\end{equation}

\begin{equation}
    F_\nu^{SN}(t)=C \frac{\exp{-\frac{t-t_0}{\tau_{fall}}}}{1+\exp{-\frac{t-t_0}{\tau_{rise}}}}+D
    \label{eq:SN}
\end{equation}

The joint approximation of all stages of the light curve without Galaxy extinction correction and host galaxy subtraction is shown in Fig.~\ref{FullLC}. The values of the parameters after host subtraction and Galaxy extinction correction are presented in Table~\ref{optic_lc_fit_parameters}.
\begin{table*}
	\centering
	\caption{Parameters of the early afterglow, afterglow, and supernova components of optical light curve joint approximation after host subtraction and Galaxy extinction correction.}
	\label{optic_lc_fit_parameters}
	\begin{tabular}{p{0.07\textwidth}p{0.07\textwidth}p{0.07\textwidth}p{0.07\textwidth}p{0.04\textwidth}p{0.07\textwidth}p{0.07\textwidth}p{0.07\textwidth}p{0.07\textwidth}p{0.03\textwidth}p{0.08\textwidth}}
        \hline
	    $A \times 10^{-5}$ (Jy) & $\omega_1$ & $x_c \times 10^{-3}$ & $B \times 10^{-6}$ (Jy) & $\alpha$ & $C \times 10^{-6}$ (Jy) & $t_0$ (days) & $\tau_{fall}$ (days) & $\tau_{rise}$ (days) & D (Jy) & $\chi^2/d.o.f.$\\
		\hline
	    $1.14 \pm 0.05$ & $0.56 \pm 0.03$ & $5.4 \pm 0.1$ & $3.0 \pm 0.3$ & 0.77* & $5.2 \pm 3.8$ & $10.2 \pm 5.1$ & $9.5 \pm 7.7$ & $1.6 \pm 1.3$ & 0* & 1.9 / 72\\
		\hline
        \multicolumn{11}{l}{* -- Parameter was fixed.}
	\end{tabular}
\end{table*}

By employing the Bazin function~\citep{Bazin2011}, which characterizes the SN component (Eq.~\ref{eq:SN}), it becomes possible to determine the parameters defining the SN peak from the joint fit. This is the time from the onset of the burst to the maximum of the SN light curve in the observer's frame and the absolute magnitude at the maximum of the supernova's brightness, corrected for Galaxy extinction.

Using the effective wavelength from~\citet{Fukugita95}, observer-frame $R$ corresponds to rest-frame $634.9 / (1+z) = 445.2$ nm. The effective wavelength of the V filter is 547.7 nm. The difference between $R_{rest}$ and V$_{obs}$ is 102 nm. To compare with other known SNe, we would need to use the I filter, because it would be closest to the V filter in the rest-frame. However, since no observations were conducted in the I filter, we assume R$_{obs}$ $\approx$ V$_{rest}$ for SN/GRB's parameters comparison.

\begin{table}
	\centering
	\caption{The parameters of the SN~201015A estimated in this work.}
	\label{SN_parameters}
	\begin{tabular}{lcccc}
		\hline
        $M_V$ (mag) & $T_{max}$ (days) & Type & $v_{phot}$ (km s$^{-1}$) & $T_{bb}$ (K)\\
        \hline
        $-19.45_{-0.47}^{+0.85}$ & $8.54 \pm 1.48$ & Ic & $1.1_{-0.2}^{+2.0}\times10^4$ & $\sim 5000$\\        
        \hline
	\end{tabular}
\end{table}

The obtained values of the time of maximum brightness after GRB trigger $T_{max}=8.45$$\pm$1.48~days (in a rest frame) and absolute magnitude $-19.45_{-0.47}^{+0.85}$ mag at maximum allow to compare it with other studied SN/GRBs cases. A comparison reveals that the SN~201015A exhibits the earliest peak among the described SN/GRB cases~\citep[e.g.][]{Belkin2020} and is in the middle of absolute magnitude distribution. 

We can also estimate the absolute magnitude at the SN maximum using the results of numerical simulations of the afterglow with the \texttt{afterglowpy} described in Section~\ref{afterglowpy}. Using the afterglow contribution at $\sim$14.2 days after the trigger, we calculate it to be $-19.53_{-0.20}^{+0.87}$ mag. This is comparable to what was obtained by approximating the entire light curve with a multicomponent function and is also in consistency with what was presented by~\cite{Patel2023}.

\begin{figure*}
	\includegraphics[width=0.9\textwidth]{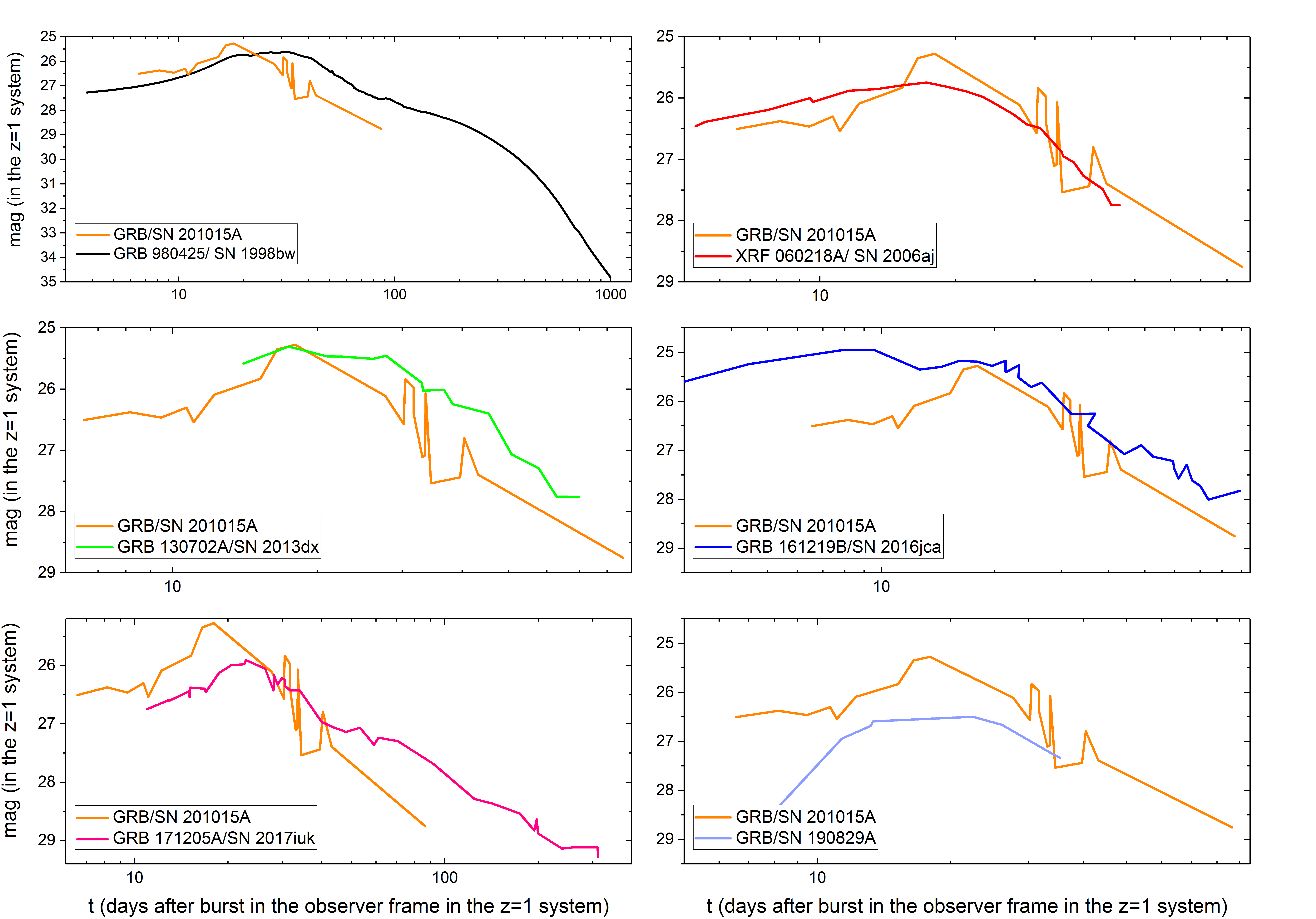}
    \caption{The light curve of the supernova associated with the GRB~201015A compared to the known light curves of SN/GRB cases (GRB~980425/SN~1998bw:~\citet{Clocchiatti2011}; GRB~060218A/SN~2006aj:~\citet{Ferrero2006}; GRB~130702A/SN~2013dx:~\citet{Volnova2017}; GRB~161219B/SN~2016jca:~\citet{Cano2017}; GRB~171205A/SN~2017iuk:~\citet{Volnova2023}; GRB~190829A:\citet{Pian2023}). Light curves are shown after recalculation them to z = 1. The afterglow and the host galaxy contribution were subtracted. All data were corrected for Galactic extinction~\citep{Schlafly2011} and host galaxy extinction, where it was determined. As can be seen, the light curves of the SN~2016jca and the SN~2006aj have the most similarity with the light curve of the SN~201015A.}
    \label{SN_LC_Comparison}
\end{figure*}

A comparison of the SN~201015A with various broad-lined Ic SN associated with GRB is shown in Figure~\ref{SN_LC_Comparison}. The comparison was carried out under the assumption SNe were observed from the redshift z=1 to compare them in the same frame. The decay rate of the SN~201015A most closely resembles that of SN~2016jca associated with GRB 161219B~\citep{Cano2017}. The closest overall light curve coincidence is seen for SN~2006aj associated with XRF~060218~\citep{Ferrero2006}, particularly in the SN's dome, suggesting that their luminosity at the SN's maximum should be approximately equal.

\subsubsection{Spectrum}
\label{SpectrumSection}
We used both MODS-1 and -2 in dual-grating mode (grisms G400L and G670L) providing a wavelength coverage from 3200--9500~\AA\ and a slit mask with a width of 1 arcsec. The total integration time was 8$\times$900 seconds.
The MODS data were reduced at the Italian LBT Spectroscopic Reduction Center \footnote{
\href{http://www.iasf-milano.inaf.it/software}{http://www.iasf-milano.inaf.it/software}
} using scripts optimised for LBT data adopting the standard procedure
for long-slit spectroscopy with bias subtraction, flat-fielding, bad-pixel correction, sky subtraction, and cosmic-rays contamination. Wavelength calibration (in the air) was obtained using the spectra of Hg, Ar, Xe, and Kr lamps providing an accuracy of $\sim$0.08 \AA~ over the whole spectral range. Relative flux calibration was derived from the observations of spectrophotometric standard stars.

All the spectra have been corrected for the Galactic extinction (A$_V$ = 0.93). The low S/N spectrum shows a peak around 540 nm (rest frame) and is consistent with type Ic-BL supernova spectra around maximum light. The spectroscopic observations concur with the photometric discovery of the supernova initially proposed in~\cite{gcn29033}.

To obtain the photospheric velocity and identify the lines, modeling was carried out using the SYNOW code~\citep[e.g.,][]{Parrent2010}, which was previously employed to describe the spectra of both ordinary SNe and objects associated with GRBs. This code can be characterised by several assumptions, such as the spherically symmetric expansion of ejecta, the emission of light from a sharp photosphere (with lower velocity in the model), and the physical modeling of line profiles in which optical depths decrease exponentially with an e-folding velocity $v_e$ ($\tau \propto e^{-v/v_e}$).

The slope of the central part of the spectrum (4000--6500 \AA) was used to fit a black-body continuum with the temperature $T_{bb} = 5000^{+3000}_{-1000}$ K. We excluded the edges of the spectrum, $<4000$ \AA\ and $>6500$ \AA, from the fitting process, as they can be affected by the presence of other lines (e.g., CaII in the blue part) or instrumental problems.

It was possible to identify only the prominent lines of FeII (S/N$\sim$2.4), which are usually observed in SN/GRB spectra. The slope within the range of 4800 \AA\ to 5200 \AA\ can be used for fitting the FeII blend and estimating the photospheric velocity. The observed spectral detail can be described by the presence of FeII expanding with velocities in the range from 5000 km s$^{-1}$ up to 30000 km s$^{-1}$. However, the best model shows $v_{ph} = 11000$ km s$^{-1}$, $v_e = 20000$ km s$^{-1}$, and a temperature of 5000 K for this spectrum of SN~201015A.

\begin{figure}
  \includegraphics[width=\columnwidth]{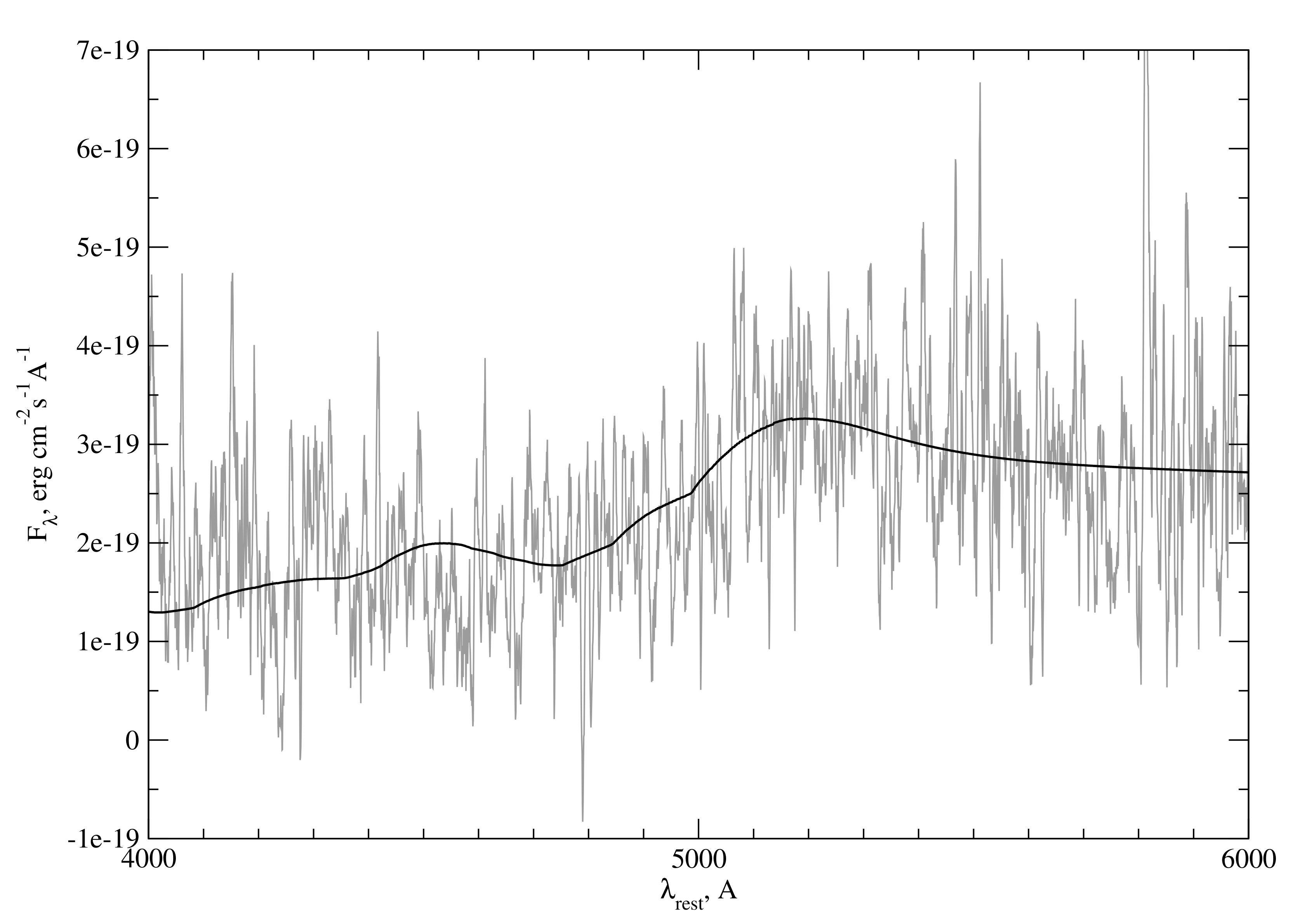}
    \caption{The resulting approximation of the SN~201015A spectrum using FeII.}
    \label{spectra_fit_fits}
\end{figure}




Fig.~\ref{spectra_fit_fits} shows the observed spectrum and the model used to approximate it.

\begin{figure}
	\includegraphics[width=\columnwidth]{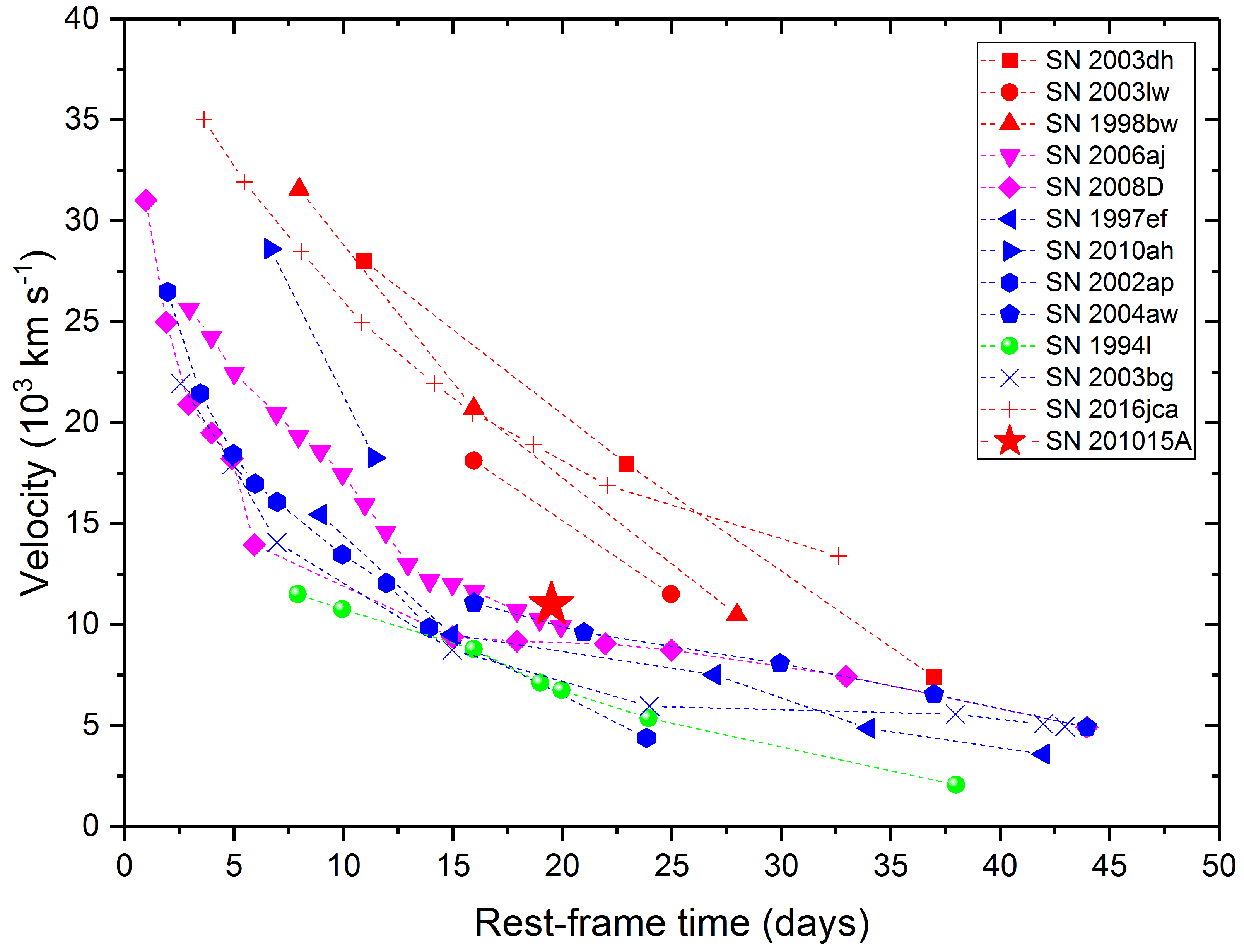}
    \caption{The photospheric velocity evolution with a rest-frame time given for 13 SNe. The red points are GRB-SN cases. The pink points are SNe associated with X-ray flashes (XRFs). The blue points are SNe unrelated to any of the mentioned high-energy events. SN 1994I is a normal SN Ic. References: ~\citet{pian2006,klose2019,ashall2019}. 
    }
    \label{PhotVelGraph}
\end{figure}

As a result of spectrum modeling, with a focus on the iron lines, the most satisfactory photospheric velocity was estimated to be 11000 km s$^{-1}$. It might be compared with certain photospheric velocities for supernovae of various types. Fig.~\ref{PhotVelGraph} depicts the evolution of photospheric velocity over time for GRB-associated SNe, SNe with X-ray flashes, and SNe with no relation to any of the transient events. Even taking into account large errors, it can be seen that the photospheric velocity for the SN~201015A still fits well with the fact that GRB-SNe have consistently higher photospheric velocities than standard events for the corresponding epoch.

\subsection{Host galaxy}
\label{host_galaxy}
For the first time, the host of the GRB 201015A was identified in the several epochs of the stacked image during the SN’s maximum stage (see Fig.~\ref{nearobjects}). These observations were conducted with the 1.5-meter telescope AZT-22 of the Maidanak Observatory on October 26, November 4 and 6. Total exposure was 61x180 seconds. The pixel size (0.268$^{\prime\prime}$/pixel) and FWHM ($\sim$0.8$^{\prime\prime}$) made it possible both to detect the source at coordinates consistent with those suggested by other research groups and to identify an object that is a candidate for the host galaxy. No sources within the coordinates of the proposed host were found in the catalogues (SDSS,  PanSTARRS, Legacy Survey). The most likely explanation is that the source's magnitude is fainter than the limiting magnitude of the images used to compile the catalogues.

The high quality of the stacked image enabled the determination of the source's offset relative to the host galaxy which is $\sim$1.5$^{\prime\prime}$. Assuming standard cosmology the scale would be 5.635 kpc$/\prime\prime$, equivalent to a projected distance of 8.453 kpc.

In addition, late-time observations with the 1.5-meter telescope AZT-20 of the Assy-Turgen observatory on January 9, 2021, were carried out. The total exposure was 160$\times$60 seconds. As a result of photometry and mathematical flux subtraction of the neighboring objects (see Section~\ref{optical_afterglow}), the host galaxy level was determined to be $24.5$ mag in the r$^\prime$ filter.

Assuming that the surface distribution of galaxies adheres to a Poisson distribution, we can calculate the probability that a given galaxy is not associated with the GRB~201015A~\citep[e.g.][]{Bloom2002}. With the given offset and the host's magnitude in the R filter after taking into account the Galaxy extinction, this probability is equal to 0.018. Consequently, it can be concluded that with a probability of 98.2\% this galaxy is the host galaxy for the GRB~201015A.




\begin{table}
	\centering
	\caption{GRB 201015A host galaxy information.}
	\label{Host_parameters}
	\begin{tabular}{lccc}
		\hline
        RA(J2000) & DEC(J2000) & $m_{r^{\prime}}^1$ (mag) & $M_V$ (mag)\\
        \hline
        23:37:16.3110 & +53:24:55.340 & $24.5$ & $-17.40$\\        
        \hline
		\multicolumn{4}{l}{$^{1}$Not corrected for foreground or rest-frame extinction.}
	\end{tabular}
\end{table}

\section{Conclusions}
\label{conclusion}

It's mostly quite challenging to spot a supernova associated with a gamma-ray burst, mainly because there are certain factors at play~\citep[e.g. distant source, bright host/afterglow,][]{Belkin2023}. At present, only about  50 SN/GRB cases were confirmed as a result of photometric observations, and there are an additional 27 GRBs where spectroscopic evidence has revealed supernovae (with significant overlap between these two groups). Moreover, a detailed light curve, and even rarer, multicolour light curves, have only been constructed for a handful of such supernovae. However, it's crucial to emphasize that discovering a SN/GRB and studying its emissions can significantly enhance and advance our understanding of the GRB phenomenon. This underscores the importance of organizing quasi-continuous follow-up observations for each nearby (low redshift) GRB detected, with the primary goal of searching for its associated supernova and potentially expanding our sample of such events.

As a result of an organised observational campaign of the GRB~201015A (z=0.426) covering its prompt emission phase and over the course of approximately 85 days following the burst, the supernova was discovered from a light curve of a source and subsequent spectroscopic observations allowed to confirm the presence of the supernova and determine the time of maximum brightness after GRB trigger $T_{max}=8.45$$\pm$1.48~days (in a rest frame) and absolute magnitude $-19.45_{-0.47}^{+0.85}$ mag at maximum.

Through obtained photometric observations, crucial parameters of the SN~201015A were extracted. It is the absolute magnitude of the SN's light curve maximum and the time of the peak from the GRB's detection in the rest frame. The comparison of these parameters with other known SN/GRB cases revealed that SN~201015A stands out as the earliest one, reaching peak luminosity in just $\sim$8.54 days after the burst detection while the median values of $T_{max}$ across all SN/GRB's sample is about 14.2 days.

As a result of numerical modeling with the \texttt{afterglowpy}~\citep{Ryan2020}, estimates of the main properties of the GRB jet and CSM were obtained ($E_k=2.0^{+3.8}_{-1.35}\times10^{53}$ erg, $n=2\times10^{-3}$ cm$^{-3}$, $\theta_j =0.30^{+0.08}_{-0.12}$, $\theta_v=0.31^{+0.30}_{-0.21}$), as well as an estimate of the afterglow component to be subtracted from the SN emission was determined. Using it, the absolute magnitude of the SN's light curve maximum was calculated and it coincides within the error with what was obtained by approximation of the light curve with a multicomponent function defined by Eq.~\ref{eq:OT}.

A phenomenological comparison was conducted between the light curve of GRB~201015A and well-studied SN/GRBs light curves. In order to do this all light curves were recalculated to the z=1 system. Notably, the decay pattern of the SN~201015A bears a resemblance to that of SN~2016jca/GRB~161219B~\citep{Cano2017}. However, when considering the overall behaviour, SN~201015A's light curve closely mirrors that of SN~2006aj, which, in turn, is associated with the XRF~060218~\citep{Ferrero2006}.


Spectroscopic observations, on the other hand, enabled the identification of the flux excess on the $\sim$28.8 day as a SN Ic-BL~\citep{gcn29306}. Using the SYNOW code~\citep[e.g.][]{Parrent2010}, the modeling of the obtained spectrum was carried out, determining the temperature of the blackbody component ($T_{bb}\sim$5000 K) and the photospheric velocity ($v_{phot}\sim$11000 km s$^{-1}$) on $\sim$16.5 days after the maximum of the light curve of the SN~201015A. The estimated value of the photospheric velocity does not contradict the distribution for this parameter determined for other SN/GRBs (Fig.~\ref{PhotVelGraph}). The value of photospheric velocity itself aligns most closely with the photospheric velocity of the SN~2006aj associated with the XRF~060218A for the respective epoch. Nevertheless, the large dispersion of values does not allow to draw reliable conclusions based on this.

The magnitude of the proposed host galaxy for GRB~201015A was determined through a mathematical flux subtraction procedure. Given the presence of multiple objects in close proximity to the host (see Fig.~\ref{nearobjects}), straightforward photometry is quite a challenging process. Instead, we employed aperture photometry, encompassing all nearby objects within the aperture radius. Subsequently, we performed flux subtraction to isolate the host galaxy's contribution. This method gave a host galaxy flux value equivalent to 24.5 mag in the r$^\prime$ filter. We consider the galaxy as the host galaxy for GRB~201015A with a 98.2$\%$ probability.

In addition to examining the optical counterpart, an analysis of the prompt emission phase (gamma-rays) of the GRB~201015A was conducted. Publicly available data from the \textit{Fermi}-GBM experiment were used. The source remains undetectable above 50 keV due to its soft spectrum~\citep[e.g.][]{Zhang2023}. To classify the source, the method suggested in \citet{min20a} was applied, which relies on the $ E_\text{p,i} $ -- $ E_\text{iso} $ correlation. The source's position on the $ E_\text{p,i} $ -- $ E_\text{iso} $ and $ T_\text{90,i} $ -- $ EH $ diagrams, as well as the value of the parameter $EH$=0.49 classify GRB~201015A as a type II GRB.

\section*{Acknowledgements}
SB, ASP, PYM, NSP, and AAV are grateful to the Russian Science Foundation (project no. 23-12-00220) for their partial support of the data reduction, analysis of data, and modeling. We also thank Sergei Blinnikov for useful discussion and suggestions regarding the article.

AR acknowledges support from the INAF project Premiale Supporto Arizona \& Italia.
Spectroscopic data were obtained thanks to the program IT-2019B-018 (P.I. A. Rossi).
The LBT is an international collaboration among institutions in the United States, Italy, and Germany. LBT Corporation partners are The University of Arizona on behalf of the Arizona Board of Regents; Istituto Nazionale di Astrofisica, Italy; LBT Beteiligungsgesellschaft, Germany, representing the Max-Planck Society, The Leibniz Institute for Astrophysics Potsdam, and Heidelberg University; The Ohio State University, representing OSU, University of Notre Dame, University of Minnesota and University of Virginia.

GS acknowledges the support by the State of Hesse within the Research Cluster ELEMENTS (Project ID 500/10.006).

EVK is grateful to the Ministry of Science and Higher Education of Russian Federation for financial support (of the work); The AZT-33IK telescope is a part of Center for Common Use «Angara».

VK and IVR were funded by the Aerospace Committee of the Ministry of Digital Development, Innovations and Aerospace Industry of the Republic of Kazakhstan (Grant No. BR11265408)

This work made use of data supplied by the UK Swift Science Data Centre at the University of Leicester.

\section*{Data Availability}
The data presented in the paper would be sent on reasonable request to the corresponding author.



\bibliographystyle{mnras}
\bibliography{example} 




\appendix
\newpage
\section{The log of the radio observations}
\begin{table*}
	\centering
	\caption{Log and results of the radio observations of the GRB~201015A}
	\label{radio_obs_log}
	\begin{tabular}{lcccc}
		\hline
	    UT Date & UT time & t-T$_0$ (d)$^{1}$ & Frequency (GHz) & Flux (Jy) \\
		\hline
		2020-10-16 & 19:02-22:41 & 0.91770 & 36.8 & 0.0034$\pm$0.0009 \\
		2020-10-17 & 19:14-21:54 & 1.90555 & 36.8 & 0.0038$\pm$0.0012 \\
		2020-10-18 & 20:12-23:14 & 2.95347 & 36.8 & 0.0042$\pm$0.0013 \\
		2020-10-19 & 21:08-23:54 & 3.98680 & 36.8 & 0.0039$\pm$0.0011 \\
		2020-10-20 & 21:31-23:35 & 4.98819 & 36.8 & 0.0034$\pm$0.0011 \\
		2020-10-21 & 21:03-23:03 & 5.96736 & 36.8 & 0.0027$\pm$0.0012 \\
		2020-10-22 & 22:13-23:44 & 7.00590 & 36.8 & 0.0018$\pm$0.0012 \\
		2020-10-23 & 21:21-23:14 & 7.97743 & 36.8 & 0.0011$\pm$0.0013 \\
		\hline
		\multicolumn{5}{l}{$^{1}$ -- The mid-exposure time relative to the trigger time (T$_0$=2459138.45153935 JD).}
	\end{tabular}
\end{table*}

\newpage
\section{The log of the optical observations}
\begin{table*}
	\centering
	\caption{GRB~201015A optical observation log. Magnitudes are in the AB (g$^\prime$, g, r, r$^\prime$, i$^\prime$, i, z$^\prime$, z) and Vega (R) systems, not corrected for foreground or rest-frame extinction}
	\label{obs_log}
	\begin{tabular}{lcccccc}
		\hline
	    UT Date & t-T$_0$ (d)$^{1}$ & Filter & Magnitude (mag) & Exposure (s) & Telescope & Source \\
		\hline
		15 Oct 2020 & 6.99E-04 & r$^\prime$ & 17.49$_{-0.20}^{+0.20}$ & 20 & FRAM & \citet{Ror23}\\
		15 Oct 2020 & 9.99E-04 & r$^\prime$ & 17.71$_{-0.23}^{+0.23}$ & 20 & FRAM & \citet{Ror23}\\
		15 Oct 2020 & 0.00113 & r & 18.00$_{-0.05}^{+0.05}$ & 3$\times$40 & NEXT & this work\\
		15 Oct 2020 & 0.00132 & r$^\prime$ & 17.61$_{-0.21}^{+0.21}$ & 20 & FRAM & \citet{Ror23}\\
		15 Oct 2020 & 0.00162 & r$^\prime$ & 17.38$_{-0.18}^{+0.18}$ & 20 & FRAM & \citet{Ror23}\\
		15 Oct 2020 & 0.00193 & r$^\prime$ & 16.87$_{-0.11}^{+0.11}$ & 20 & FRAM & \citet{Ror23}\\
		15 Oct 2020 & 0.00223 & r$^\prime$ & 16.74$_{-0.10}^{+0.10}$ & 20 & FRAM & \citet{Ror23}\\
		15 Oct 2020 & 0.00252 & r$^\prime$ & 16.68$_{-0.10}^{+0.10}$ & 20 & FRAM & \citet{Ror23}\\
		15 Oct 2020 & 0.00262 & r & 16.97$_{-0.03}^{+0.03}$ & 4$\times$60 & NEXT & this work \\
		15 Oct 2020 & 0.00282 & r$^\prime$ & 16.73$_{-0.10}^{+0.10}$ & 20 & FRAM & \citet{Ror23}\\
		15 Oct 2020 & 0.00312 & r$^\prime$ & 16.73$_{-0.10}^{+0.10}$ & 20 & FRAM & \citet{Ror23}\\
		15 Oct 2020 & 0.00343 & r$^\prime$ & 16.97$_{-0.12}^{+0.12}$ & 20 & FRAM & \citet{Ror23}\\
		15 Oct 2020 & 0.00373 & r$^\prime$ & 16.75$_{-0.10}^{+0.10}$ & 20 & FRAM & \citet{Ror23}\\
		15 Oct 2020 & 0.00403 & r$^\prime$ & 16.64$_{-0.09}^{+0.09}$ & 20 & FRAM & \citet{Ror23}\\
		15 Oct 2020 & 0.00433 & r$^\prime$ & 16.90$_{-0.12}^{+0.12}$ & 20 & FRAM & \citet{Ror23}\\
		15 Oct 2020 & 0.00463 & r$^\prime$ & 16.68$_{-0.10}^{+0.10}$ & 20 & FRAM & \citet{Ror23}\\
		15 Oct 2020 & 0.00493 & r$^\prime$ & 16.79$_{-0.11}^{+0.11}$ & 20 & FRAM & \citet{Ror23}\\
		15 Oct 2020 & 0.00523 & r$^\prime$ & 16.77$_{-0.11}^{+0.11}$ & 20 & FRAM & \citet{Ror23}\\
		15 Oct 2020 & 0.00553 & r$^\prime$ & 16.75$_{-0.10}^{+0.10}$ & 20 & FRAM & \citet{Ror23}\\
		15 Oct 2020 & 0.00583 & r$^\prime$ & 16.94$_{-0.12}^{+0.12}$ & 20 & FRAM & \citet{Ror23}\\
		15 Oct 2020 & 0.00613 & r$^\prime$ & 17.01$_{-0.13}^{+0.13}$ & 20 & FRAM & \citet{Ror23}\\
		15 Oct 2020 & 0.00643 & r$^\prime$ & 17.09$_{-0.14}^{+0.14}$ & 20 & FRAM & \citet{Ror23}\\
		15 Oct 2020 & 0.00673 & r$^\prime$ & 17.41$_{-0.09}^{+0.09}$ & 20 & FRAM & \citet{Ror23}\\
		15 Oct 2020 & 0.00793 & r$^\prime$ & 17.64$_{-0.10}^{+0.10}$ & 20 & FRAM & \citet{Ror23}\\
		15 Oct 2020 & 0.00913 & r$^\prime$ & 17.65$_{-0.09}^{+0.09}$ & 20 & FRAM & \citet{Ror23}\\
		15 Oct 2020 & 0.00924 & r & 18.15$_{-0.10}^{+0.10}$ & 12$\times$90 & NEXT & this work\\
		15 Oct 2020 & 0.01033 & r$^\prime$ & 17.95$_{-0.13}^{+0.13}$ & 20 & FRAM & \citet{Ror23}\\
		15 Oct 2020 & 0.01154 & r$^\prime$ & 18.19$_{-0.16}^{+0.16}$ & 20 & FRAM & \citet{Ror23}\\
		15 Oct 2020 & 0.01665 & R & 18.79$_{-0.12}^{+0.12}$ & 20 & FRAM & \citet{Ror23}\\
		15 Oct 2020 & 0.03989 & R & 19.88$_{-0.28}^{+0.28}$ & 20 & FRAM & \citet{Ror23}\\
		15 Oct 2020 & 0.09584 & White & 20.27$_{-0.08}^{+0.08}$ & 6$\times$120 & ZTSh & this work\\
		15 Oct 2020 & 0.10470 & White & 20.51$_{-0.08}^{+0.08}$ & 6$\times$120 & ZTSh & this work\\
		15 Oct 2020 & 0.11358 & White & 20.54$_{-0.07}^{+0.07}$ & 6$\times$120 & ZTSh & this work\\
		15 Oct 2020 & 0.12243 & White & 20.67$_{-0.08}^{+0.08}$ & 6$\times$120 & ZTSh & this work\\
		15 Oct 2020 & 0.13059 & White & 20.77$_{-0.08}^{+0.08}$ & 5$\times$120 & ZTSh & this work\\
		15 Oct 2020 & 0.13799 & White & 20.67$_{-0.08}^{+0.08}$ & 5$\times$120 & ZTSh & this work\\
		15 Oct 2020 & 0.14598 & R & 20.70$_{-0.14}^{+0.15}$ & 120 & ZTSh & this work\\
		15 Oct 2020 & 0.15143 & R & 20.94$_{-0.25}^{+0.27}$ & 120 & ZTSh & this work\\
		15 Oct 2020 & 0.15581 & R & 21.15$_{-0.24}^{+0.26}$ & 120 & ZTSh & this work\\
		15 Oct 2020 & 0.16017 & R & 21.19$_{-0.33}^{+0.36}$ & 120 & ZTSh & this work\\
		15 Oct 2020 & 0.16382 & R & 20.97$_{-0.22}^{+0.24}$ & 120 & ZTSh & this work\\
		15 Oct 2020 & 0.16675 & R & 21.19$_{-0.32}^{+0.35}$ & 120 & ZTSh & this work\\
		16 Oct 2020 & 0.50417 & R & 22.21$_{-0.24}^{+0.31}$ & 9$\times$300 & 2.16-m & this work\\
		16 Oct 2020 & 0.64075 & R & 22.31$_{-0.75}^{+0.91}$ & 30$\times$120 & AZT-33IK & this work\\
		16 Oct 2020 & 0.68081 & r$^\prime$ & 22.91$_{-0.30}^{+0.39}$ & 38$\times$60 & AZT-20 & this work\\
		16 Oct 2020 & 0.68796 & g$^\prime$ & 23.23$_{-0.49}^{+0.37}$ & 36$\times$60 & AZT-20 & this work\\
		16 Oct 2020 & 0.69640 & i$^\prime$ & 21.93$_{-0.43}^{+0.32}$ & 38$\times$60 & AZT-20 & this work\\
		16 Oct 2020 & 0.71278 & R & 22.59$_{-0.48}^{+0.52}$ & 108$\times$30 & AC-32 & this work\\
		16 Oct 2020 & 0.80396 & R & 22.38$_{-0.11}^{+0.11}$ & 87$\times$120 & ZTSh & this work\\
		16 Oct 2020 & 0.89426 & Rc & 22.86$_{-0.29}^{+0.32}$ & 12$\times$300 & Zeiss-1000 (S) & this work\\
		17 Oct 2020 & 1.73782 & r$^\prime$ & 23.92$_{-0.66}^{+1.05}$ & 75$\times$60 & AZT-20 & this work\\
		17 Oct 2020 & 1.78383 & Rc & 23.67$_{-0.31}^{+0.39}$ & 12$\times$300 & Zeiss-1000 (S) & this work\\
		18 Oct 2020 & 2.70303 & r$^\prime$ & 24.24$_{-0.80}^{+0.88}$ & 53$\times$60 & AZT-20 & this work\\
		18 Oct 2020 & 2.73227 & R & 23.36$_{-0.69}^{+0.76}$ & 34$\times$120 & AZT-33IK & this work\\
		18 Oct 2020 & 2.74040 & R & 23.69$_{-0.56}^{+0.60}$ & 198$\times$30 & AC-32 & this work\\
		19 Oct 2020 & 3.74775 & r$^\prime$ & 24.04$_{-0.76}^{+0.76}$ & 97$\times$60 & AZT-20 & this work\\
		19 Oct 2020 & 3.89597 & R & 23.58$_{-0.29}^{+0.41}$ & 8$\times$300 & AZT-22 & this work\\
		20 Oct 2020 & 4.66042 & R & 22.97$_{-0.20}^{+0.31}$ & 12$\times$300 & AZT-22 & this work\\
		20 Oct 2020 & 4.78555 & r$^\prime$ & 24.20$_{-0.97}^{+1.02}$ & 77$\times$60 & AZT-20 & this work\\
		21 Oct 2020 & 5.81591 & r$^\prime$ & 23.50$_{-0.48}^{+0.84}$ & 71$\times$60 & AZT-20 & this work\\
		\hline
	\end{tabular}
\end{table*}

\begin{table*}
	\centering
	\begin{tabular}{lcccccc}
		\hline
	    UT Date & t-T$_0$ (d)$^{1}$ & Filter & Magnitude, mag & Exposure (s) & Telescope & Reference\\
		\hline
		22 Oct 2020 & 6.75409 & R & 23.33$_{-0.54}^{+0.71}$ & 68$\times$60 & AZT-33IK & this work\\
		23 Oct 2020 & 7.61789 & R & 23.02$_{-0.80}^{+0.95}$ & 69$\times$60 & AZT-33IK & this work\\
		23 Oct 2020 & 7.88468 & g & >23.6 & 180 & GTC & this work\\
		23 Oct 2020 & 7.88765 & r & 23.62$_{-0.55}^{+1.12}$ & 180 & GTC & this work\\
		23 Oct 2020 & 7.89064 & i & 22.26$_{-0.26}^{+0.32}$ & 180 & GTC & this work\\
		23 Oct 2020 & 7.89430 & z & 21.40$_{-0.21}^{+0.26}$ & 300 & GTC & this work\\
		24 Oct 2020 & 8.70510 & R & 22.95$_{-0.39}^{+0.62}$ & 21$\times$60 & Zeiss-1000 (T) & this work\\
		26 Oct 2020 & 10.84952 & R & 22.86$_{-0.34}^{+0.47}$ & 19$\times$180 & AZT-22 & this work\\
		28 Oct 2020 & 11.75488 & R & 22.59$_{-0.78}^{+0.96}$ & 90$\times$60 & AZT-33IK & this work\\
		28 Oct 2020 & 12.81643 & R & 22.55$_{-0.28}^{+0.36}$ & 27$\times$180 & AZT-22 & this work\\
		04 Nov 2020 & 19.73793 & R & 23.06$_{-0.28}^{+0.37}$ & 27$\times$180 & AZT-22 & this work\\
		06 Nov 2020 & 21.55624 & R & 23.27$_{-0.86}^{+1.03}$ & 48$\times$60 & AZT-33IK & this work\\
		06 Nov 2020 & 21.62762 & R & 23.08$_{-0.42}^{+0.67}$ & 30$\times$150 & Zeiss-1000 (T) & this work\\
		06 Nov 2020 & 21.72694 & R & 22.93$_{-0.35}^{+0.49}$ & 15$\times$240 & AZT-22 & this work\\
		07 Nov 2020 & 22.61255 & R & 23.21$_{-0.54}^{+0.67}$ & 30$\times$120 & AZT-33IK & this work\\
		07 Nov 2020 & 22.61255 & r$^\prime$ & 23.52$_{-0.66}^{+1.34}$ & 56$\times$60 & AZT-20 & this work\\
		08 Nov 2020 & 23.57483 & R & 23.46$_{-0.67}^{+0.73}$ & 30$\times$120 & AZT-33IK & this work\\
		08 Nov 2020 & 23.87368 & Rc & 23.45$_{-0.36}^{+0.42}$ & 11$\times$300 & Zeiss-1000 (S) & this work\\
		08 Nov 2020 & 23.94498 & r & 23.57$_{-0.51}^{+0.54}$ & 540 & GTC & this work\\
		08 Nov 2020 & 23.95339 & z & 22.16$_{-0.47}^{+0.81}$ & 600 & GTC & this work\\
		09 Nov 2020 & 24.56936 & R & 23.61$_{-0.71}^{+0.87}$ & 30$\times$120 & AZT-33IK & this work\\
		13 Nov 2020 & 28.22051 & r & 24.26$_{-0.69}^{+0.81}$ & 14$\times$120 & LBT & this work\\
		13 Nov 2020 & 28.22054 & g & >25 & 14$\times$120 & LBT & this work\\
		13 Nov 2020 & 28.80000 & r & 23.89$_{-0.20}^{+0.20}$ & 14$\times$120 & LBT & this work\\
		15 Nov 2020 & 30.77597 & R & 23.56$_{-0.31}^{+0.42}$ & 15$\times$240 & AZT-22 & this work\\
		13 Dec 2020 & 58.87929 & r & 24.17$_{-0.79}^{+1.17}$ & 540 & GTC & this work\\
		13 Dec 2020 & 58.88762 & z & 22.21$_{-0.60}^{+1.26}$ & 600 & GTC & this work\\
		16 Dec 2020 & 61.64333 & R & 23.82$_{-0.60}^{+1.32}$ & 69$\times$120 & AZT-33IK & this work\\
		16 Dec 2020 & 61.67941 & R & 23.93$_{-0.48}^{+0.81}$ & 6$\times$240 & AZT-22 & this work\\
		09 Jan 2021 & 85.67848 & r$^\prime$ & 24.45$_{-0.99}^{+n/d}$ & 160$\times$60 & AZT-20 & this work\\
		\hline
		\multicolumn{5}{l}{$^{1}$ -- The mid-exposure time relative to the trigger time (T$_0$=2459138.45153935 (JD)).}
	\end{tabular}
\end{table*}


\bsp	
\label{lastpage}
\end{document}